# Taking a Look into Execute-Only Memory

Marc Schink
*Fraunhofer Institute AISEC*
marc.schink@aisec.fraunhofer.de

Johannes Obermaier
*Fraunhofer Institute AISEC*
johannes.obermaier@aisec.fraunhofer.de

## Abstract

The development process of microcontroller firmware often involves multiple parties. In such a scenario, the Intellectual Property (IP) is not protected against adversarial developers which have unrestricted access to the firmware binary. For this reason, microcontroller manufacturers integrate eXecute-Only Memory (XOM) which shall prevent an unauthorized read-out of third-party firmware during development. The concept allows execution of code but disallows any read access to it. Our security analysis shows that this concept is insufficient for firmware protection due to the use of shared resources such as the CPU and SRAM. We present a method to infer instructions from observed state transitions in shared hardware. We demonstrate our method via an automatic recovery of protected firmware. We successfully performed experiments on devices from different manufacturers to confirm the practicability of our attack. Our research also reveals implementation flaws in some of the analyzed devices which enables an adversary to bypass the read-out restrictions. Altogether, the paper shows the insufficient security of the XOM concept as well as several implementations.

## 1 Introduction

Embedded systems and microcontrollers in particular became popular for a number of application fields like robotics, transportation and medicine. Concepts such as the Internet of Things (IoT) even increase their pervasiveness in more areas in industry as well as in consumer products. As a consequence, such devices comprise increasingly complex software, making the contained Intellectual Property (IP) more valuable to adversaries. Common protection techniques prevent firmware read-outs by disabling the debug interface of a microcontroller. Such mechanisms are applicable against outside attackers that are not involved in the firmware development process.

However, complex software is often not developed by a single company but built upon software provided by other companies, so called *third-party software*. In a multi-party development process, there is no protection against adversarial developers and they have unrestricted access to the binary code. Therefore, every developer is able to reverse engineer or copy the software, and thereby pose a threat to the contained IP. Since developers require debug interface access and privileged code execution on a device, common firmware protection mechanisms that disable the debug features are not applicable. For that reason, a firmware protection technique against adversarial developers in multi-party development scenarios is required. In such a scenario, the firmware is deployed and secured on the device before handing it over to the next developer. As a consequence, developers need physical access to the device such that their firmware can be developed and deployed in a trusted environment. In summary, a security concept for multi-party development needs to protect firmware against an adversarial developer with the following capabilities: physical access to the microcontroller, its integrated debug interface as well as arbitrary and privileged code execution. Hardware modification capabilities are not considered because they may be detectable by customers or other involved parties of the development process.

An approach that enables firmware protection for embedded devices and microcontrollers in multi-party development environments has been described by ARM [24]. This approach is based on eXecute-Only Memory (XOM) which allows solely the execution of code but prevents any read or write access to it. This feature is present in various microcontrollers from different manufacturers. Also, it has gained attention through support in toolchains and the embedded systems community [5, 17]. While the article by ARM notes that adversaries might be able to partially guess protected code by observing changes in CPU registers an SRAM content, the article also states that such an attack requires *significant effort*. However, no details nor evidence from real embedded systems are provided.

For this reason, this paper presents the following contributions:

- An analysis of XOM as protection technique against unauthorized code read-out.

- An evaluation and discovery of flaws in the XOM concept of several ARM Cortex-M based microcontrollers.
- A general procedure to automatically recover firmware protected by XOM.
- A practical evaluation of code recovery attacks on devices from different manufacturers.
- The discovery and exploitation of further implementation flaws that allow reading XOM protected code.

## 1.1 Related Work

Code protection mechanisms have been investigated by several researchers and conceptual as well as implementation flaws were discovered. Obermaier and Tatschner presented three (non-)invasive attacks against the firmware read-out protection of STM32F0 devices [11]. Goodspeed and Francillon demonstrated an attack which leverages the bootloader implementation of MSP430 microcontrollers to dump the protected flash memory content [4]. They describe how to recover specific instructions from its unknown firmware based on whether the device performs a reset after a code injection attack. The found instructions are then used as *gadgets* for a Return-Oriented Programming (ROP) attack to circumvent the read-out protection. Bittau et al. describe how to recover specific instructions to mount a ROP attack on unknown code [3]. A similar attack which targets the Intel Software Guard Extensions (SGX) was presented by Lee et al. [6]. The software in the secure enclave can be executed but the binary is unknown to the attacker, similar to XOM. The authors use an exception signal to guess instructions inside the enclave. Again, only very specific instructions are recovered for a subsequent ROP attack.

These results are only partially applicable to XOM protected memory. Despite the situation of a read-out protected memory is similar, entirely recovering protected code has never been shown to be feasible.

## 2 Execute-Only Memory

In this section, we start with a description of the general idea behind eXecute-Only Memory (XOM) as firmware protection technique and its application for multi-party development scenarios. Subsequently, we explain the conceptual weakness that arises from it, and how this can be exploited to circumvent the protection.

### 2.1 General Concept

The fundamental idea behind XOM-based software protection mechanisms is to ensure confidentiality and integrity by preventing read-outs and manipulations of the respective code, and thereby protect the contained IP.

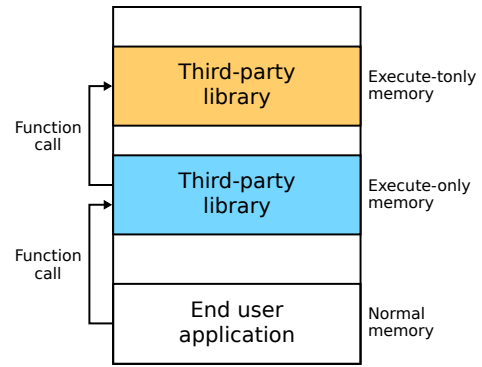

Figure 1: Flash memory with end user application and two third-party firmware components marked as execute-only.

Some approaches exist to integrate this concept into desktop and server systems [22, 23]. Since an adversary with physical access can easily probe and read-out their main memory or data storage, the CPU is the only trusted hardware component. The encrypted code is stored in memory and is decrypted and executed on-demand within the CPU.

Small embedded systems and microcontrollers in particular are usually based on a single chip incorporating CPU and memory. Thus, probing and data read-out of the memory is impossible without invasive hardware attacks. Since the firmware can only be accessed by components within the device, such as the processor, XOM is implemented by partially restricting modifications and read-outs of the flash memory while allowing instruction fetches. Encryption of the code is not necessary. This enables protection against unintended or malicious code read-outs, for example, data leakage bugs due to missing bound checks. Also, it provides protection against malicious or accidental firmware modifications which increases the safety of a system.

In a multi-party firmware development scenario, XOM allows developers to deploy their software on a microcontroller and subsequently mark it as *execute-only*. After that, the microcontroller including the deployed software can be utilized by other developers but the software is protected against malicious read-outs. Figure 1 illustrates an example use case where the flash memory contains two libraries together with an end user application. The libraries are deployed on the device by third-party software providers. Since both libraries are placed inside the XOM, they can be utilized but are protected against read-outs and modifications by the developer of the end user application.

### 2.2 Conceptual Weakness

The combination of XOM as a firmware protection technique and the extensive capabilities of an adversarial developer results in a *conceptual* weakness.

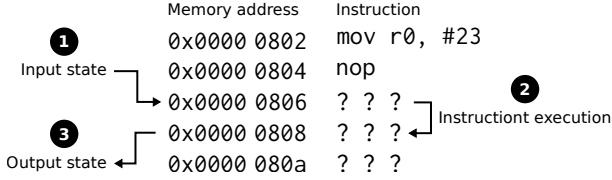

Figure 2: Execution flow to gather system state changes.

In contrast to a Trusted Execution Environment (TEE) that provides isolated code execution, XOM provides only separated flash memory regions but shares all other resources, such as CPU and SRAM. Even though the read-out restrictions of XOM prevent direct access to the firmware, the execution on shared resources leaks information about the protected code. Once an instruction is fetched, whether execute-only or regular code, the same resources are used for execution, and thereby the protection vanishes. Due to its extensive capabilities, an adversarial developer is able to execute code in XOM and observe its effect on these resources. The adversary exploits the CPU as so called *instruction oracle* to infer information about the protected instructions. In the following, we call a *system state* the accessible part of these shared resources. A system state includes, but is not limited to, the CPU registers as well as the SRAM content. Based on the *changes* in the system state, an adversary is able to infer information about executed instructions. This is what we call *instruction recovery*.

In Figure 2, an exemplary illustration of an XOM disassembly is depicted, unknown instructions are represented by question marks. The figure shows the three necessary steps to gather system state changes from an unknown instruction to be recovered. We denote such an instruction as *target instruction*. In this example, it is located at 0x0000 0806. The system states before and after executing a target instruction are referred to as *input* and *output system state*, respectively. The steps to gather system state changes caused by the target instruction are the following:

1. Setup of the system input state for the target instruction.
2. Execution of the target instruction inside the XOM.
3. Read-out of the modified system state.

The modification (1) and read-out (3) of the system state require only the common capabilities of a developer. In order to obtain system state changes caused only by the target instruction (2), it must be executed individually. An adversary can achieve this by utilizing the microcontroller's debug capabilities and its single-stepping feature.

A more general way to gather system state changes for instruction recovery is possible by utilizing an interrupt to execute only the target instruction. For example, a timed interrupt can be used to limit code execution to the target instruction only. Once the CPU reaches the interrupt handler, its state can be obtained. Even though the exception entry modifies the Program Counter (pc) and other registers, the state *after* the execution of the target instruction can be restored from the stack. Hence, we call this approach *interrupt-driven* instruction recovery. Its advantage is that it does not require any debug capabilities but only privileged code execution on the target device.

## 2.3 Instruction Recovery

Using either the single-stepping or interrupt-driven approach, we are able to obtain system state changes from a target instruction, and thus everything we need for instruction recovery. Based on the setup depicted in Figure 2, we describe the basic idea of the recovery process with a single set of input and output state, as listed in Table 1. For simplicity, the system states only comprise two registers. Considering the difference of the pc between input and output state, we deduce that the target instructions must be 16 bit wide. The modification of r0 does not clearly identify the target instruction. It may result from an add r0, #0x1f, but also other instructions such as ldr and mov are possible. However, we can already exclude branch instructions because they never modify r0. In order to fully recover the target instruction, additional input states and their corresponding output states are necessary.

Table 1: Exemplary input and output system state for an unknown target instruction.

| Register | Input state | Output state |
| --- | --- | --- |
| r0 | 0x23 | 0x42 |
| pc | 0x0000 0806 | 0x0000 0808 |

Instruction recovery requires observable system state changes which are not available under some conditions. An article about XOM already lists various situations for which no change in system state is observable [24]:

1. Store instruction operating on a write-only location
2. Data processing instruction producing the same result as the previous destination register value
3. Compare or test instruction with unchanged result flags
4. Conditional branch that is not taken
5. Execution of a memory barrier or hint instruction

However, since ARM Cortex-M devices comprise a load-store architecture, every data processing and memory instruction operates on the core registers of the CPU, and thus can be manipulated by an adversarial developer. As a result, the ambiguities (1) to (4) can be resolved by choosing appropriate input system states. The input system state for a particular instruction is chosen such that it propagates its unique characteristics into the system output state, and thereby can be distinguished from other instructions. A single instruction may be executed multiple times with different input states in order to be fully identified. Some instructions (5) do not necessarily depend on an accessible part of the system state nor alter it in an observable manner. Nonetheless, including information, such as a side-channel for the execution time of an instruction, might be an additional feature to make them recoverable.

## 3 Device Analysis

In the following, we analyze XOM implementations for multi-party firmware development from different manufacturers. We focus on how to obtain system state changes that leak information about protected instructions. All examined devices are listed in Table 6 in Appendix A.1.

### 3.1 STM32 Devices

A prominent XOM implementation is called Proprietary Code Read-Out Protection (PCROP) and was developed by STMicroelectronics. Various STM32 microcontroller families comprise this feature, ranging from the low-power STM32L0 series to the high-performance STM32H7 series [14, 15].

The STM32Lx and STM32F4 devices do not restrict any debug capabilities such as single-stepping when executing protected code. Changes in the system state, such as CPU registers and SRAM content, are observable. Hence, an adversary can gather system state changes of protected instructions and can base code recovery attacks thereon.

In contrast, the STM32F7 and STM32H7 devices incorporate the core debug functionality into the XOM security concept. The CPU cannot be halted via the debugger while executing protected code. Attempts to perform single-stepping inside the XOM are ignored. However, the CPU can be halted whenever code outside the secured memory is executed. For example, when the processor executes a function or an interrupt handler implemented outside the secure code region. Interrupts will still be handled and allow therefore to halt the CPU whenever desired. For that reason, target instructions inside the XOM can be executed with arbitrary system states and the corresponding output states are observable with the interrupt-driven recovery approach.

### 3.2 Tiva C and MSP432 Devices

Texas Instruments implements XOM as firmware protection technique in various device families [18, 19, 21]. The Tiva C and MSP432E4 series do not restrict debug capabilities during execution of protected code, and thus single-stepping is possible. On the contrary, the MSP432P4 series restricts the debug features and prevents to halt the CPU while executing protected code. It disconnects the debug interface on single-stepping attempts inside a protected code region. Also, SRAM access during execution of protected code is blocked to avoid data read-outs via the debug interface [20]. Nonetheless, whether during single-stepping on Tiva C and MSP432E4 devices or once the CPU returns from execution of secured memory on MSP432P4 microcontrollers, system state changes caused by protected instructions are observable to an adversary.

A unique feature of the MSP432P4 series allows the user to perform load operations inside the secured memory. Thereby, data placed within the XOM is accessible by the protected code. This feature can be enabled during initial configuration and additionally needs to be unlocked by the protected code itself. Load operations from secured memory are disabled by default and are not recommended when utilizing XOM for IP protection according to the data sheet [19].

### 3.3 Kinetis Devices

The Kinetis microcontroller series by NXP Semiconductors integrates a XOM-based firmware protection feature in various sub-families [8–10].

The corresponding application note for these devices indicates potential security issues when debug access is enabled and consequently limits the use-case of the firmware protection. The same document proposes to disable the debug interface as mitigation against potential attacks [7]. Nevertheless, to the authors' knowledge, there is no publicly available security analysis of this firmware protection technique. For that reason, the affected devices are included in this research. Similar to most of the analyzed devices, single-stepping inside the XOM is possible to obtain system state changes and with that also code recovery attacks.

This microcontroller family allows `pc`-relative load instructions inside protected code regions. This enables protected code to access constant values, so called *literal pools*, located in XOM. These instructions can always be used and do not need to be explicitly enabled [7].

## 4 Code Recovery Attack

In the following section, we describe in more detail how the instruction recovery process works and how the necessary input states are designed. Based on that, we present a setup to recover protected code and evaluate it on two devices from different manufacturers. Subsequently, a proof of concept setup for interrupt-driven code recovery is presented. We finish this section by elaborating the limitations for code recovery attacks and discuss possible countermeasures.

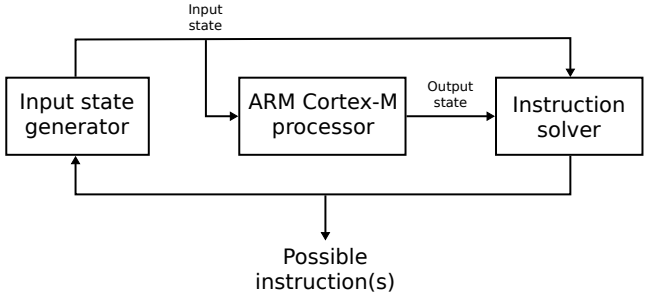

Figure 3: Basic components and flow of the instruction recovery process.

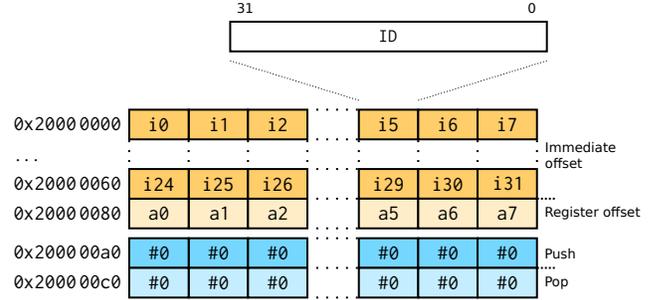

Figure 4: SRAM content of input state for memory instructions.

## 4.1 Recovery Process

In Section 2.3, we explained how to recover instructions from a high level perspective. In what follows, we describe a process that allows instruction recovery in an automated fashion.

The three steps outlined below are executed as long as a target instruction is not fully determined and further input states are available. The flow diagram of this process is illustrated in Figure 3 and is made up of the following steps:

1. **Input state generation:** The *input state generator* creates system states based on all possible candidates for the target instruction. A well chosen input state for the initial iteration helps reducing the number of possible candidates early in the recovery process, thereby decreasing the required iterations. For the second and following iterations, the *input state generator* takes the result of the previous iterations into account to generate the next system input state.

2. **Instruction execution:** After a suitable system state is generated, the CPU executes the, still unknown, target instruction and produces the system output state. In this step, the CPU is exploited as *instruction oracle* that provides information about the target instruction.

3. **Instruction solving:** During the final step, the *instruction solver* rules out instructions based on the input and output state of the system. The solving procedure is divided into the following steps:

    - **Pre-check:** This step is used to filter out possible instructions by simple rules based on the input and output state. The goal is to cancel out instructions in an early stage before performing the more complex and time consuming *enumeration* and *verification* steps. For example, only some instructions are able to update the processor status flags, ruling out the others.
    - **Enumeration:** This step is performed only once for a target instruction and enumerates all possible instructions based on the initial input and output state.
    - **Verification:** During this step, each possible instruction is evaluated on the input state and the result is checked against the observed output state of the target instruction. Instructions leading to incorrect results are discarded and will no longer be considered.

After this process, there may be multiple possibilities for a target instruction because some properties cannot be inferred from the system state. We elaborate on this and other limitations of the presented recovery process later in Section 4.5.

## 4.2 Design of System Input States

In what follows, we describe how input states are designed in order to identify instructions with their individual properties.

The input states are tailored for the ARMv6-M architecture and its instruction set. We chose this architecture because it is the common basis for all the examined devices. We exemplarily cover several prevalent instruction groups and describe how to build input states to recover them.

### 4.2.1 Regular Load and Store Instructions

In general, a memory instruction comprises a destination register `rt`, an address base register `rn` and either an immediate offset value `imm` or an offset register `rm`. For example, `ldr rt, [rn, #imm]` denotes a load instruction with immediate offset and `ldr rt, [rn, rm]` with register offset. The ARMv6 architecture specifies three base register types: the lower registers `r0` to `r7`, the Stack Pointer (`sp`) and the Program Counter (`pc`). In this section, we cover only the first one. The other two types will be described subsequently. The architecture supports memory operations with different properties like element sizes and signedness. In the following, we explain how to recover 32 bit memory instructions.

For code recovery, the SRAM content and the register values are chosen such that executing a memory instruction will consecutively reveal the base and destination register as well as offset parameters. Figure 4 illustrates the SRAM *layout* and the corresponding addresses. The first part of the memory is dedicated to instructions with immediate offset. It comprises 32 words, denoted as `i0` to `i31`, corresponding to each possible immediate offset value. Similarly, the second part targets instructions with register offset and therefore contains eight words, denoted as `a0` to `a7`. Each word corresponds to a possible offset register `r0` to `r7`.

The SRAM *content* is chosen such that each word contains a 32 bit unique identifier (ID), as depicted in the upper region of Figure 4. When a load instruction is being executed, a memory word appears in a register, thereby revealing its destination register as well as the source address and thereby its underlying offset. All register values are chosen to differ from the memory words, thus, loading a memory word will *always* modify the destination register's value. The same applies when storing a register value to memory.

Table 2: Register assignment of input state for memory instructions.

| Register | Value | Register | Value |
|---|---|---|---|
| r0 | 0x80 | r5 | 0x2000 0000 |
| r1 | 0x2000 0000 | r6 | 0x8c |
| r2 | 0x84 | r7 | 0x2000 0000 |
| r3 | 0x2000 0000 | sp | 0x2000 00c0 |
| r4 | 0x88 | | |

However, no register assignment exists that can guarantee that solely valid SRAM addresses are accessed. The reason is that we do not know the base register of the memory instructions in advance and might thereby access an invalid address which causes a hard fault. This being the case, we use a straight-forward assignment for `r0` to `r7` as listed in Table 2. Half of the registers contain the SRAM offset address, while the others contain offset values pointing to the region dedicated to instructions with *register offset*. This has shown to be an adequate trade-off between covering immediate/register offset instructions and avoiding hard faults.

The recovery process starts with the aforementioned input state and is illustrated in the following example. By executing the unknown instruction, we observe that the memory value of `i5` was written to `r2`. This implies that a load instruction with `r2` as destination register was executed. The memory word `i5` is located at address `0x2000 0014`. We infer the base address as `0x2000 0000` and an immediate offset of 20 bytes. There are no other instructions that are able to load this memory word for the given input state. Since the base address is present in multiple registers, the base register `rn` is either `r1`, `r3`, `r5` or `r7`. At this point, we partially recovered the instruction as `ldr r2, [rn, #20]`. In the next iteration of the recovery process, we individually increment the values of the possible base registers by 0, 4, 8 and 12 such that they point to *different* memory addresses. After executing the instruction again, we observe that the value of `i7` was written to `r2`. This memory word is located at address `0x2000 001c`. Given the immediate offset of 20 bytes, we can infer `r5` as the base register with the address `0x2000 0008` since `0x2000 001c = 0x2000 0008 + 20`. Finally, the memory instruction is fully recovered as `ldr r2, [r5, #20]`.

Store instructions are handled in a similar manner. The SRAM is checked for any changes caused by store instructions. The address of the stored word enables us to infer the addressing mode including the base register and offset. The value written to the memory reveals the source register of the store instruction.

Depending on the target memory instruction, multiple iterations are required to resolve ambiguities. This is automatically handled by our framework which selects the next input states accordingly. Some instructions might even lead to hard fault exceptions due to accessing invalid memory addresses. For example, when `r1` and `r3` are used as base and offset register, respectively. Nevertheless, any information, including the occurrence of hard faults, facilitates narrowing down the set of possible memory instructions.

#### 4.2.2 Push and Pop Instructions

These memory instructions use the `sp` as base register and have a set of registers to load or store 32 bit values. Because they have a fixed base register, we only need to distinguish between both instructions and recover their register set. For example, the register set of `push {r0, r4}` contains `r0` and `r4`. A register set can contain up to eight registers from any of the registers `r0` to `r7` and either the Link Register (`lr`) or `pc` for a `push` and `pop` instruction, respectively.

Table 3: Register assignment of input state for push and pop instructions.

| Register | Value | Register | Value |
|---|---|---|---|
| r0 | 0xf0 | r7 | 0xf7 |
| ... | ... | sp | 0x2000 00c0 |
| r6 | 0xf6 | lr | 0xf8 |

Whether an instruction is a `push` or `pop` is distinguished by the modification of the `sp`, it decreases for a `push` and increases for a `pop` instruction. The remaining property that needs to be recovered is the register set. For this, we use the lower half of the SRAM as depicted in Figure 4. The memory layout is split into two parts for `push` and `pop`, respectively. The memory words in this region do not require a special content but must be distinguishable from the register values.

This ensures that an instruction *always* causes observable changes in the registers or the memory content. For this input state, we use a zero value for all memory words. The register assignment for the recovery process is listed in Table 3. It ensures that both instructions operate on their designated memory region. In order to identify the register set of a `push` instruction, we use unique register values for all possible registers `r0` to `r7` and `lr`. Once a `push` instruction is executed, the unique register values are written to the SRAM and identify the corresponding registers. For example, a `push {r4, r7}`, writes the register values `0xf4` and `0xf7` into the first two words of the dedicated memory region. The register set of a `pop` instruction can clearly be identified by the modified register values. For example, a `pop {r0, r7}` loads a zero value into the registers `r0` and `r7`. The same holds for the special case where the `pc` is included in register set.

### 4.2.3 PC-Relative Load Instructions

These instructions are similar to regular load instructions with immediate offset but use the `pc` as base register. They are used to load constant values placed in code memory. Due to the read-out restriction of XOM, literal data cannot be located within the secured memory region, and thus the immediate value relative to the `pc` cannot be determined. For that reason, the recovery process represents `pc`-relative loads by the literal data instead of the immediate offset. This input state is also necessary for devices that do not support memory loads from protected memory. The reason is that `pc`-relative loads inside the XOM may access unprotected memory regions nearby.

The input state for regular memory instructions already narrows the target instruction down to the group of load instructions. The only ambiguity that may arise is that literal data is equal to one of the SRAM words in Figure 4. In that case, a `pc`-relative and regular load cannot be distinguished from each other. We resolve this ambiguity with a special memory address assigned to all possible base registers of regular load instructions. The address is chosen such that it is not mapped in the system address space and any read causes a fault. For example, a possible address is `0x8000 0000`. If the target instruction is a regular load, a fault occurs because every base register contains an illegal memory address. Otherwise, the literal data is loaded because `pc`-relative loads are independent of the assigned base registers. This way, we are able to recover `pc`-relative load instructions.

### 4.2.4 Branch Instructions

There are multiple instructions that can manipulate the `pc` and thereby perform a branch. We start with a description of how to recover branches that use a register to specify the target instruction address: `blx`, `bx`, `mov` and `pop`. In order to identify this register, we assign unique values to all possible registers. Table 4 shows the corresponding register assignment. The Least Significant Bit (LSB) of every register value is cleared to ensure that `blx`, `bx` and `pop` instructions change the Thumb state, making them distinguishable from a `mov` instruction. Since the `pop` instruction is already covered by the procedure to recover `push` and `pop` instructions, we use the same SRAM and `sp` configuration as described before. The `blx`, `bx` and `pop` instructions are clearly identifiable by how they modify the `lr` and `sp`.

Table 4: Register assignment of input state for branch instructions.

| Register | Value | Register | Value |
|---|---|---|---|
| r0 | 0x2000 00e0 | r7 | 0x2000 0060 |
| ... | ... | ... | ... |
| r6 | 0x2000 0050 | lr | 0x2000 00d0 |

The other group of branch instructions use an immediate value that encodes the branch address as difference between the current and target instruction address, hence called *pc-relative branches*. The immediate value is recovered by calculating the difference of the `pc` between the output and input state. To establish a clear distinction between the former group of branch instructions, we use the SRAM address `0x2000 0000` as base value for the register assignment, as shown in Table 4. The distinction arises from the fact that XOM-protected code is executed in a different address space, usually at its beginning. This way, `pc`-relative branches with their limited range cannot intersect with the target address of the former branch instruction group. As a consequence, we only need to distinguish between the two `pc`-relative branches `b` and `bl`. Since only the latter modifies the `lr`, both instructions can be unambiguously recovered. The cleared LSB of the `lr` ensures that a modification is always observable.

Conditional branches, such as `beq`, are special `b` instructions and require additional input states to be recovered. We use multiple input states that modify the corresponding flags in the Application Program Status Register (APSR) to enforce all possible branching conditions. With the help of the instruction timing, we are able to determine whether a branch was taken or not taken and thereby recover the branch condition. We are able to recover every branch instruction and, except for conditional ones, need only a single input state and iteration in the recovery process.

## 4.3 Code Recovery via Single-Stepping

In the previous sections, we showed how to gather system state changes from a device, how to automatically recover instructions and how the corresponding input states are designed. With that, we have all the building blocks necessary for a code recovery attack on a microcontroller.

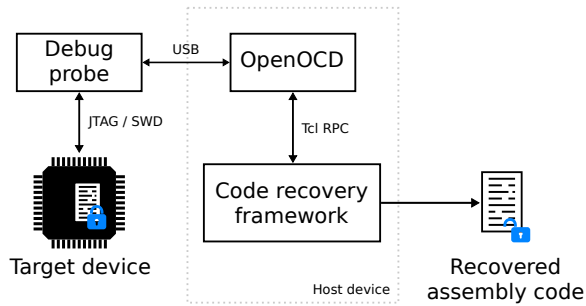

Figure 5: Setup for code recovery utilizing the single-stepping feature of the target device.

We chose the Kinetis KV11 and STM32L0 microcontrollers as target because they comprise the ARMv6-M architecture which our system input states are designed for. Also, they allow single-stepping inside the XOM which allows us to gather system state changes by utilizing the integrated debug features. We employ Python to implement a code recovery framework based on the process explained in Section 4.1 and input states similar to those described in Section 4.2.

Figure 5 depicts the code recovery setup with all its components and their interconnections. We use the integrated debug probes of the development boards to establish a connection between the host device, that runs the code recovery framework, and the target device. On the host device, Open On-Chip Debugger (OpenOCD) is used to interface the debug probes. The code recovery framework uses OpenOCD to orchestrate the recovery process.

Table 5: Evaluation of code recovery via single-stepping on Kinetis KV11 and STM32L0 microcontrollers.

| Target device | Flash memory | Instruction recovery time | Flash recovery time |
|---|---|---|---|
| Kinetis KV11 | 128 KiB | 2.33 s | $\sim 42$ h |
| STM32L0 | 192 KiB | 1.64 s | $\sim 45$ h |

The code coverage and the recovery speed are the key indicators whether the presented attack is practicable and thereby poses a threat to firmware protected by XOM. In order to determine the recovery speed, we used the filter function `arm_conv_partial_opt_q15()` of the CMSIS-DSP library as benchmark. Due to its broad coverage of the ARMv6-M instruction set, we chose this filter in favor of others, including one that is used in an example scenario [12] for the PCROP. Table 5 shows the results of the performance evaluation for both microcontrollers. On average, it takes 2.33 s and 1.64 s to recover an instruction from a Kinetis KV11 and STM32L0, respectively. The variation may be due to the different target devices or debug probes. The driver implementation of the debug probes in OpenOCD may also have an influence. However, we did no further investigation on this topic. Given the flash memory size of both microcontrollers, we estimated the time it takes to recovery the whole code. For this we assume the worst-case where the entire flash memory is protected and only 16 bit wide instructions are used. As a result, the entire flash memory can be recovered in less than two days on both devices. We consider these results as a reasonable amount of time for an adversary to perform code recovery attacks in a multi-party development scenario. In reality, an attack will even succeed much faster as only a part of the flash memory is usually marked as execute-only. In terms of code coverage, we are able to fully recover the protected function with a few limitations described in Section 4.5

In summary, we are able to recover the code from secured memory within a reasonable time, thus demonstrating the practicality of code recovery attacks against XOM-protected firmware and its contained IP.

### 4.4 Interrupt-Driven Code Recovery

The previously described attack uses the single-stepping feature of the debug interface. Since the STM32F/H7 and MSP432P4 devices prevent this with a more advanced security concept, the previous approach is not applicable to obtain system state changes.

We use the interrupt-driven code recovery approach as explained in Section 2.3 to circumvent this security concept. It needs malicious code on the target device to apply and read-out system states as well as execute the target instructions. This code could be injected and executed on the target device via a vulnerability or a regular firmware update mechanisms. We implemented a proof of concept that abuses the SysTick timer to generate interrupts after the target instruction is executed. The accuracy is achieved by clocking the timer with the core clock frequency. Therewith, we are able to obtain the system state changes caused by the target instruction. The malicious code is executed in SRAM and requires no modification of the flash memory. We use UART as interface between the host and the target device which allows us to mount an attack even if the debug interface is disabled. The setup is depicted in Figure 6. Since we are able to obtain system state changes of protected instructions, a similar procedure as described before can be applied to carry out a code recovery attack. We modified the recovery framework for a proof of concept and evaluated this approach in a semi-automatic fashion for different groups of instructions on the STM32F/H7 and MSP432P4. The limitations of this approach are described in Section 4.5.

This setup has also been successfully tested on devices that do not restrict single-stepping inside the XOM, such as the Kinetis KV11 device family. This enables code recovery attacks even on devices with a disabled debug interface.

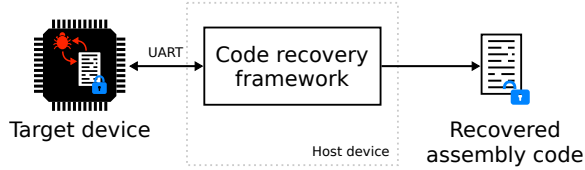

Figure 6: Setup for code recovery utilizing the interrupt-driven approach.

## 4.5 Limitations

Instruction recovery based on observing the results of different inputs has the fundamental limitation that some instructions cannot be identified unambiguously because they are identical from an algorithmic perspective.

This applies to instructions with commutative properties. For example, the instructions `add r0, r1, r2` and `add r0, r2, r1` are equal and indistinguishable. Moreover, some instructions can be represented by different binary encodings that are equal from an algorithmic point of view. This limitation is negligible as none of these affect the correct *functionality* of the extracted code. We counter these ambiguities in the recovery framework by a post-processing step which chooses the resulting instruction based on recommendations of the architecture specification [2] and heuristic rules. However, some instructions pose a limitation to the recovery process. Nevertheless, they are comparatively rare and are needed only for special purposes. Table 7 in Appendix A.2 lists the instruction groups of the ARMv6-M architecture and to what extent they can be recovered.

The interrupt-driven code recovery approach that we use to overcome the disabled single-stepping feature introduces further limitations. At first, the search space increases because the expiration value of the SysTick timer needs to be adapted to the execution time of the unknown instruction. It may be necessary to execute the unknown instruction multiple times with increasing timer values until the instruction is executed. Apart from that, the execution of only a single instruction cannot be guaranteed on the STM32F/H7 series. Multiple instructions may be folded and executed during a single step which further increases the search space. Finally, instructions that influence the interrupt behavior, such as `cps`, are problematic for this approach since it uses an interrupt to obtain the output system state. Special care must be taken for these instructions such that the protected code is not executed in its entirety if the interrupt is not triggered. Altogether, the limitations are comparatively small and do not prevent extraction of IP.

## 4.6 Mitigation

Since the discovered issues are founded within the devices' immutable hardware, there exists no fix but only mitigations. Software based countermeasures are mostly ineffective since adversaries can command the control flow. A debug probe as well as malicious code can both take over full system control, thus, there is no possibility to detect or prevent an attack from within the protected code because every instruction is executed in isolation. As a consequence, there are no countermeasures except disabling the debug interface and restricting any execution of privileged code on the device. However, this prevents firmware development in multi-party environments, thereby nullifying the advantage of XOM.

From a hardware development point of view, an appropriate countermeasure against the presented attack would be a TEE, such as ARM TrustZone, that executes protected firmware in a separate execution environment.

## 5 Exploiting Implementation Flaws

Besides the conceptual weakness of the evaluated XOM-based protection mechanisms, our research uncovered flaws in their hardware implementation. The missing isolation between protected and unprotected code allows adversaries to exploit these vulnerabilities and bypass the read-out restrictions. In this section, we describe our findings and how they can be developed into a vulnerability. The results are listed in Table 6 in Appendix A.1 as *read-out* vulnerabilities.

## 5.1 STM32F7 Devices

The additional and dedicated Instruction Tightly-Coupled Memory (ITCM) bus between flash module and processor is a unique feature of the STM32F7 microcontroller. The flash module is accessible via two different address ranges starting from `0x0800 0000` and `0x0020 0000` for the AMBA High-performance Bus (AHB) and the ITCM bus, respectively [1,13,15,16]. An analysis of this additional bus revealed that *data* transfers from the flash memory are blocked, with a single exception: data transfers initiated by the AHB Access Port (AHB-AP) are not blocked correctly. As a consequence, the protected code on these devices can be read-out via the debug interface without any restrictions, thereby circumventing the security feature. We assume that every access via the AHB-AP through the ITCM bus is erroneously executed as an *instruction* fetch instead of a data transfer which is why reads are not filtered correctly.

## 5.2 Kinetis Devices

The Kinetis microcontrollers support `pc`-relative loads inside the XOM [7]. We found out that not only `pc`-relative but all load instructions inside a secured memory region can read from it. This is of particular importance because it poses a threat to the confidentiality of the contained code: regular load instructions inside the XOM can be exploited to circumvent its read-out restrictions.

Since input and output state are accessible to an attacker, regular load instructions can be executed with an arbitrary base register value pointing to the secured memory to read the code and expose it to the output state. A *gadget* is a sequence of instructions inside the secured memory suitable to extract its content. The steps are the following:

1. Setup of the system state such that the gadget is executed and reads from the protected memory.
2. The gadget is executed and loads the protected memory into its destination register.
3. A deliberate interrupt takes place to prevent execution of further instructions. This ensures that destination register value is not altered by subsequent instructions.
4. The protected content in the destination register is exposed to the system state and accessible to an adversary.

A gadget for Kinetis devices with ARM Cortex-M0+ processor consists of a single regular load instruction. Microcontrollers with ARM Cortex-M4 processor require that a certain number of non-load instructions are executed when entering the XOM and *before* a load operation is carried out. We found out that in contrast to the corresponding data sheet [7], a single preceding non-load instruction is sufficient. As a result, a regular load operation preceded by a single non-load instruction forms a suitable gadget. The code recovery process explained in Section 4 can be used to find an applicable gadget inside the secured memory. Depending on the gadget's surrounding instructions, it may be possible to relax the timing constraints for the interrupt (3). Thus, additional instructions located after and/or before the actual gadget can be executed without impeding the code extraction. We employ a similar approach as for the interrupt-driven code recovery to automatically extract protected code.

### 5.3 MSP432P4 Devices

The MSP432P4 series also allows load operations inside the secured memory, which leads to the assumption that it has the same code read-out vulnerability as the Kinetis devices.

Besides the initial configuration, this feature needs to be enabled by writing a certain key into an unlock register from within the secured memory. The very same approach as previously explained for the Kinetis devices can be utilized to execute a store instruction in a way that the key is written from within the secured code. Once unlocked, we expected to be able to perform code read-outs via a regular load instruction, similar to the approach for the Kinetis devices. However, we encountered a countermeasure that automatically locks the data access whenever code execution leaves the secured memory. As a consequence, the unlock and code read-out instructions need to be carried out without executing code outside the secured memory in between. Hence, a suitable gadget comprises a store instruction followed by a regular load instruction. To cope with the 16 bit unlock key, the store must be at least a half-word operation. Since we have to carry out both instructions without interruption, the base registers must be distinct such that both can be assigned individually before executing the gadget. The target registers must be different from the base register but apart from that we have no other constraints. The addressing mode of both instructions is not of major importance because the base register values can be adjusted to compensate potential offsets. Such gadgets are not unusual and can be found in various functions of the CMSIS library, for example. In general, there are many other instruction combinations that can be utilized as gadget. Like for the Kinetis devices, interrupt-driven code recovery can be utilized to find and automatically execute gadgets inside the secured code.

## 6 Conclusion and Outlook

In this paper, we evaluated XOM as firmware protection mechanism against adversarial developers in multi-party development scenarios. We conclude that analyzed implementations of XOM are inadequate to provide sufficient security for their intended usage. None of the analyzed devices provide sufficient protection of the firmware once an adversary has debug access or is able to execute privileged code on a microcontroller. A firmware protection technique for general multi-party development requires strict and hardware-backed isolation between different protection domains as provided by TEEs. This would have also prevented the exploitation of the hardware vulnerabilities we identified during our research.

Our methods can be extended by an implementation of input states for the ARMv7-M architecture and thereby enable full code recovery on other affected microcontrollers. Additionally, enhanced debug and trace features, such as the Data Watch and Trace (DWT) component, could be taken into account to accelerate the recovery process. Finally, there may be other devices and even architectures utilizing XOM as protection mechanism for multi-party firmware development that shall be analyzed for comparable issues.

## 7 Responsible Disclosure

As part of a responsible disclosure process, we informed the security teams of all affected manufacturers about our findings. Technical and detailed information were provided more than 120 days prior to the publication of this paper. On the request of the manufacturers and due to the variety of deployed devices utilizing XOM, we decided to not publish the code recovery framework nor other exploits at this point in time.

We thank Texas Instruments and NXP Semiconductors for their quick response and constructive discussion during the entire process.

## A Appendix

### A.1 Analyzed devices

Table 6 lists all evaluated devices, whether the integrated single-stepping feature is available in XOM, their vulnerabilities and associated CVE numbers. The *code recovery* and *read-out* vulnerabilities refer to the attacks described in Section 4 and Section 5, respectively.

Table 6: Analyzed microcontrollers with their XOM-related properties, vulnerabilities and CVE numbers.

| Manufacturer | Device family | Device under test | Single-stepping | Vulnerability | CVE |
|---|---|---|---|---|---|
| STMicroelectronics | STM32L0<br>STM32L1<br>STM32L4<br>STM32F4 | STM32L072CZ (Z)<br>STM32L152RC (V)<br>STM32L432KCU6 (Z)<br>STM32F429ZI (Y) | Yes | Code recovery | CVE-2019-14236<br>CVE-2019-14238 |
| | STM32F7<br>STM32H7 | STM32F722ZET6 (A)<br>STM32H743ZIT6 (Y) | No | Code recovery[b] / read-out[c]<br>Code recovery[b] | |
| NXP Semiconductors | K8x<br>KV1x<br>KV3x | MK82FN256VLL15 (1)<br>MKV11Z128VLF7 (0)<br>MKV31F512VLL12 (1) | Yes | Code recovery / read-out[a] | CVE-2019-14237<br>CVE-2019-14239 |
| Texas Instruments | TM4C12x | TM4C123GH6PM (B2)<br>TM4C1294NCPDT (A2) | Yes | Code recovery | CVE-2018-18056 |
| | MSP432 | MSP432E401Y (A2)<br>MSP432P401R (D) | Yes<br>No | Code recovery<br>Code recovery / read-out[a] | |

[a] Code extraction gadget (data access for MSP432P4) required.
[b] Limited code recovery, see Section 4.5.
[c] Debug interface access required.

### A.2 Code Recovery

Table 7 shows all instruction groups of the ARMv6-M architecture and to what extent they can be recovered. Every group marked with ✓ is fully recoverable, whereas (✓) indicates that at least one instruction is not fully recoverable. Ambiguities that are equal from a functional perspective are not shown in detail for the sake of clarity. Instructions that cannot be recovered because they are indistinguishable form instructions with a different functional behaviour are maked with ✗ and are explicitly listed.

Table 7: Recoverability of the ARMv6-M instruction set.

| Instruction | Recoverable | Ambiguity | Comment |
|---|---|---|---|
| **Branch** | (✓) | - | All ambiguities are equal from a functional perspective. |
| **Data processing** | (✓) | - | All ambiguities are equal from a functional perspective. |
| **Shift** | (✓) | - | All ambiguities are equal from a functional perspective. |
| **Multiply** | ✓ | - | - |
| **Packing and unpacking** | ✓ | - | - |
| **Miscellaneous data-processing** | ✓ | - | - |
| **Status register access** | ✓ | - | - |
| **Load and store** | (✓) | - | All ambiguities are equal from a functional perspective. |
| **Miscellaneous** | | | |
| `dmb` | ✗ | `dsb`<br>`isb` | Rarely used instruction. For example, it is used when modifying the vector table. |

Table 7: Recoverability of the ARMv6-M instruction set.

| Instruction | Recoverable | Ambiguity | Comment |
|---|---|---|---|
| `dsb` | ✗ | `dmb` `isb` | Rarely used instruction. For example, it is used when configuring a Memory Protection Unit (MPU). |
| `isb` | ✗ | `dmb` `dsb` | Rarely used instruction. For example, it is used when configuring an instruction cache. |
| `nop` | ✗ | `mov rd, rd` `add sp, sp, #0` `sub sp, sp, #0` `yield` `sev` | - |
| `sev` | ✗ | `nop` | Rarely used instruction. For example, it is used for signaling in multi-processor devices. |
| `yield` | ✗ | `nop` | Usually not supported and implemented as `nop`. Not supported by any of the analyzed devices. |
| **Exception-generating** | | | |
| `svc #imm` | (✓) | - | The `imm` value is not recoverable but also cannot be used in XOM. |
| `bkpt #imm` | (✓) | - | Used for debug purpose only. The `imm` value is not recoverable but also cannot be used in XOM. |